# 1720 nm – 1800 nm tunable chirped pulse amplification based on thulium-doped fiber

XINYANG LIU [1,*] AND REGINA GUMENYUK [1,2]

[1]*Laboratory of photonics, Tampere University, Korkeakoulunkatu 3, Tampere 33720, Finland*
[2]*Tampere Institute for Advanced Study, Tampere University, Kalevantie 4, Tampere 33100, Finland*
*\*xinyang.liu@tuni.fi*

**Abstract:** Chirped pulse amplification (CPA) has been adopted as a commonly used methodology to obtain powerful ultrashort laser pulses since its first demonstration. However, wavelength-tunable CPA systems are rarely reported. Wavelength-tunable ultrashort and intense laser pulses are desired in various fields like nonlinear spectroscopy and optical parametric amplification. In this work, we report a 1720 nm -1800 nm tunable CPA system based on a single-mode Tm-doped fiber covering the middle wavelength band of the third biological window. The tunable CPA system delivers ultrashort pulses varied between ~300 to 500 fs depending on the central wavelength emission at the fixed repetition rate of 22.7 MHz. The maximum average power ranges from 126 mW at 1720 nm to 294 mW at 1800 nm following the gain shape of Tm-doped fiber. Considering the specific wave-length range, this tunable CPA system is highly desired for biomedical imaging, sensing and parametric amplifiers for mid-infrared light generation.

## 1. Introduction

The concept of CPA technique was originally from an approach in radar development to get high-energy exploratory radio pulse and short detecting radio pulse to reach long detection rang while having good range resolution [1,2]. After about four decades, in 1985 the same concept was applied to laser pulse amplification [3] and be-came the golden rule to get high-power ultrashort laser pulses ever since. The CPA technique has enabled generation of ultrahigh-intensity laser pulses and facilitated many light-matter interaction applications [4].

Basically, the CPA technique enables increasing laser intensity by preventing detrimental nonlinear effects generation during light interaction with propagating medium, including both Kerr and Raman nonlinearities. In the CPA method, the low-power laser pulse generated by a seed laser is stretched temporally before being amplified in order to keep low pulse peak power in an amplifier and avoid nonlinearity growth. The amplification stage usually contains several amplifiers. After being amplified, the laser pulse is recompressed in pulse compressor. In early years, the CPA technique was mostly used in Ti: Sapphire and Nd-doped solid-state laser systems [5,6]. Gradually, the approach was widely applied in fiber laser systems, covering wavelength range from 1 μm to 2 μm using Yb, Er and Tm doped active fibers [7-9]. Research on CPA technique has thoroughly covered different stages of a multi-cascade laser system. For seed laser, it mainly determines central wavelength of the system and generated spectrum width. By engineering laser cavities, wavelength tunable and du-al-wavelength lasers have been used in the CPA system to enable wavelength tuning operation and dual-wavelength operation [10,11]. Despite the dramatic progress in ultrabroadband emission of the seed laser by careful engineering of intracavity parameters [12,13], its amplification is restricted by the gain width of the amplifier medium, resulting in the requirement to develop alternative approaches such as gain-managed nonlinear broadening to achieve sub-100 fs pulses although limited in average power level [14,15]. For a stretcher stage, several kinds of wavelength-dispersive components have been employed, like passive fiber, transmissive gratings, dispersive mirrors and chirped bragger grating [3,11,16,17]. Different amplification media have been used to boost average power levels, including rear-earth-ion doped crystal, rear-earth-ion doped fiber and rear-earth-ion doped photonic crystal fiber (PCF) [16-18]. For a com-pressor stage, a Treacy-type compressor consisting of two gratings is the most used approach due to the high damage threshold and flexibility [19], while dispersive mirrors and PCF are also employed for pulse compression [16,20]. To improve the pulse quality from the CPA system, self-phase modulation-induced pulse pedestal and gain narrowing effect have been studied, and various solutions have been put forward [6,21-24].

Wavelength tunability is a crucial feature of laser sources when it comes to laser applications such as spectroscopy, metrology, microscopy, and quantum technology. In bioimaging, one tunable laser source may cover several absorption peaks of various biological chromophores [25]. In parametric amplifiers, changing the wavelength of a tunable pump laser can alter the idler signal wavelength in a relatively large range [26]. In the past, tunable CPA systems have been realized in Ti: Sapphire solid-state laser systems at around 800 nm and Yb-doped fiber laser systems at around 1 μm [11,27]. The tuning wavelength range is usually of several tens of nm. In contrast, Tm-doped fiber can provide a wide emission wavelength band from 1600 nm to 2200 nm, exceeding other commonly used optically-active-ion doped active fibers [28] and making it a good active medium laser for a widely tunable CPA system. Part of this wavelength range falls into the third biological window (1600 nm -1870 nm), which has already been proved to have a longer attenuation length in some scattering tissues compared with the first and second biological windows [25]. The relatively long wavelength can also facilitate parametric amplifier in mid-infrared light generation to get high quantum efficiency.

In this contribution, we present a widely tunable Tm-doped fiber-based CPA system operating from 1720 nm to 1800 nm. The seed laser works in a dissipative soliton regime using a dispersion-managed cavity and delivers ps laser pulses with spectrum widths ranging from 25 nm to 34 nm. A normal-dispersion fiber is used to temporally stretch the laser pulses. After two-stage amplifiers based on single-mode fibers, the maximum average output power of 126 mW is obtained at 1720 nm with 507 fs pulse duration and 264 mW average output power is obtained at 1800 nm with a pulse duration of 294 fs. To suppress amplified spontaneous emission (ASE) at longer wavelengths, wavelength-dependent fiber bending loss is introduced before amplifiers as a tunable filter. This tunable CPA system is built for a European Union Horizon 2020 project - Advanced Multimodal Photonics Laser Imaging Tool for Urothelial Diagnosis in Endoscopy (AMPLITUDE) [29].

## 2. Experiment, Results and Discussion

The setup of the tunable CPA system fully composed of only single-mode fibers is shown in Figure 1. Principal parts of the multi-cascade laser architecture are framed separately, including a seed laser, stretcher, first amplifier, second amplifier and compressor. The seed laser is a dispersion-managed cavity composed of 9.37 m single-mode fiber and 41 cm free space. 80 cm thulium-doped fiber (TDF, TmDF200, OFS) provides efficient gain for the whole wavelength tuning range while pumping by a self-made fiber-baser laser source at 1550 nm via wavelength-division multiplexer (WDM). The pumping power is varied from 0.6 to 1 W depending on the central operation wavelength. 5.7 m dispersion compensating fiber (DCF1, UHNA4, Nufern) is employed to compensate the cavity dispersion for shifting it to slightly normal region. Another fiber in the cavity is single-mode SMF28 fiber. The net dispersion of the cavity is around 0.0044 ps2. Three waveplates and an acousto-optic tunable filter (AOTF, AOTF8, AA Opto-Electronic) are accommodated in the free space part and are responsible for non-linear polarization rotation (NPR) mode-locking and wavelength tuning, respectively. The polarization-independent isolator maintains unidirectional laser operation. One fused fiber coupler with a splitting ratio of 30/70 is used to couple out 70% optical power. The next system cascade configured for pulse stretching is realized in 10 m DCF2 (DM1010-D, YOFC). The DCF2 features large normal dispersion specified as 0.1275 ps2/m - 0.2167 ps2/m at 1550 nm as it originally designed for dispersion compensation at this wavelength. However, in this setup, apart from dispersion compensation, we take advantage of the wavelength-dependent bending loss of this fiber to suppress undesired amplified spontaneous emission (ASE) at a longer wavelength. For the first amplifier stage, 70 cm TDF is used to provide gain. The length is determined by using a cutting-back method to achieve optimum amplification performance. Following the 70 cm TDF, a WDM and a fiber-pigtailed isolator (ISO) are spliced consequently to extract the residual pump and prevent back reflection, respectively. After the ISO, 40 cm DCF2 is spliced to filter the ASE at a longer wavelength generated at the Amplifier1 stage similarly to the one after the stretcher. Another piece of 70 cm TDF is used in Amplifier2 to further enhance output power. After Amplifier2, one grating pair (T-711-1650, LightSmyth Technologies) compensates for the up-chirp of the laser pulse, composing the compressor stage. The pump for TDFs in seed laser, Amplifier1 and Amplifier2 are provided by three self-made master oscillator power amplifiers (MOPA) at 1.55 μm with a maximum average power of 3 W.

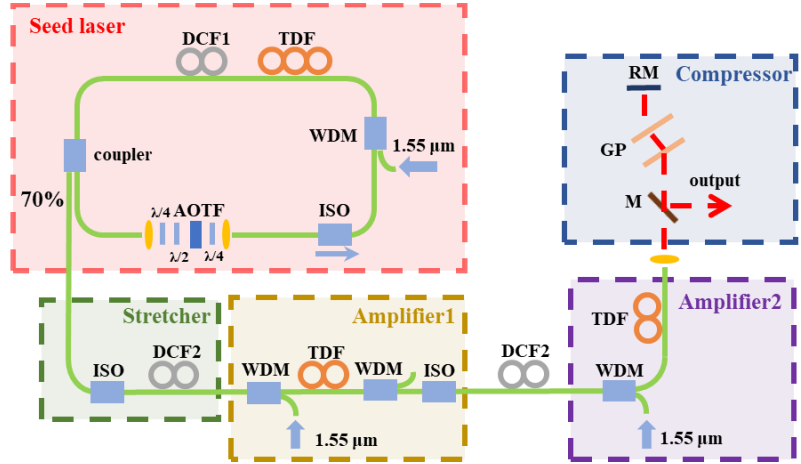

**Figure 1.** The setup of the tunable all-single-mode fiber-based CPA system. WDM: wavelength-division multiplexer; TDF: Tm-doped fiber; DCF: dispersion-compensation fiber; λ/4: quarter waveplate; λ/2: half waveplate; AOTF: acousto-optic tunable filter; ISO: isolator; GP: grating pair; RM: roof mirror.

The seed laser is at first operated in an all-anomalous-dispersion regime. After getting soliton operation, DCF1 is added to shift net cavity dispersion to slightly nor-mal and enable the laser to work in a dissipative soliton regime. The AOTF has a transmission bandwidth of around 12 nm in the operation wavelength region with a Gaussian-like transmission profile. Sweeping the frequency of the driving RF signal for AOTF from 40 MHz to 38.2 MHz, the center wavelength of AOTF can be changed from 1800 nm to 1720 nm. At 1720 nm, 1740 nm, 1760 nm, 1780 nm and 1800 nm, the pump powers for mode-locking operations are 1.06 W, 0.683 W, 0.584 W, 0.723 W and 0.683 W, respectively. All output spectra have a rectangular shape with spectral widths ranging from 25 nm to 34 nm, as shown in Figure 2 (a). The seed laser pulse durations are varied from 5.8 ps to 8.1 ps and the average output powers are from 5.45 mW to 9.3 mW, as illustrated in Figure 2 (b). One typical pulse train is shown in Figure 2 (c). Compared with our previous work [30], the seed laser performance here is optimized in terms of output power and spectral width. The variation in the output laser param-eters is caused by changes in intracavity parameters (dispersion, nonlinearity, gain and loss) within the tuning wavelength range and following it a requirement to adjust the performance of NPR to maintain stable mode-locking operation is needed.

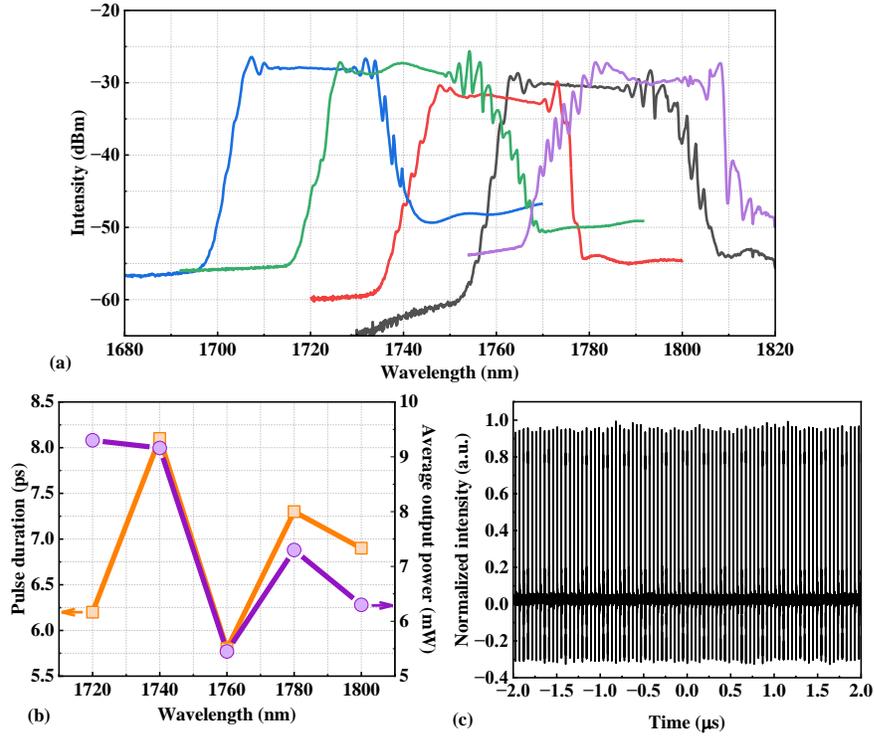

**Figure 2.** The tunable seed laser performance. (a) Optical spectra, (b) pulse durations and average output powers, (c) typical pulse train.

The laser pulses from the seed laser are subsequently stretched in 10 m DCF2 up to 72 ps at 1740 nm measured by fast photodetector with 25 GHz bandwidth (Newport 1414) and oscilloscope with 20 GHz bandwidth and 50 GS/s sampling rate (Tektronix DSA 72004). Since we aim for several hundred mW output average power in a sin-gle-mode fiber, the pulse duration is considered suitable for the following amplifica-tion. The power amplification is first done at the short wavelength edge of 1720 nm. At maximum pump power for both Amplifier1 and Amplifier2, the maximum output power after the grating pair is 126 mW with a compressed pulse duration of 507 fs, resulting in a pulse energy of 6 nJ. As this wavelength is at the short edge of the Tm-doped fiber gain profile, we note the rise of amplified spontaneous emission in the long-wavelength region at each stage of the CPA system composing by TDF (seed, am-plifier 1 and 2). Therefore, to introduce long wavelength ASE suppression, we bend the DCF2 fiber piece after the Amplifier1 with the bending diameter of ~10 cm. The DCF2 fiber features of wavelength-dependent bend-induced losses due to a small core area resulting in discrimination of waveguiding conditions for different wavelengths. Fig-ure 3 (a) illustrates the effect of introducing bending loss of DCF2 to compress ASE at longer wavelengths. With the bending loss, the ASE peak shifts from 1771 to 1758 nm with suppression by 13 dB, indicating the wavelength-dependent bending loss can be effectively used as an ASE filter in the same DCF, which can also introduce temporal stretching. This simplifies the scheme design decreasing the number of needed compo-nents for efficient operation. The suppression of long-wavelength ASE improves the spectrum extinction (the signal peak to the ASE peak ratio), showing more than 19 dB level difference. The laser operation with close to 1720 nm central wavelength experi-ences gain narrowing at the short wavelength side of the spectrum due to sharp de-creasing edge of the gain profile of Tm-doped fiber [30]. Due to this gain narrowing, the spectrum is severely reshaped with a 5-dB bandwidth of around 10 nm at the am-plification stage.

At the longest wavelength edge of the laser tunable range of 1800 nm, the DCF2 bending is released as the wavelength is close to the gain peak and ASE at shorter wavelengths is not pronounced. The spectral width at this wavelength is 27 nm with the spectrum extinction of 20 dB. The maximum output power after the grating pair is 264 mW with a compressed pulse duration of 294 fs, leading to a pulse energy of 12.7 nJ. The pedestal indicates there are some nonlinear chirp accumulations during ampli-fication, which can be further optimized by introducing stronger pulse stretching. However, the total power concentrated in the pedestal is about 15 % only.

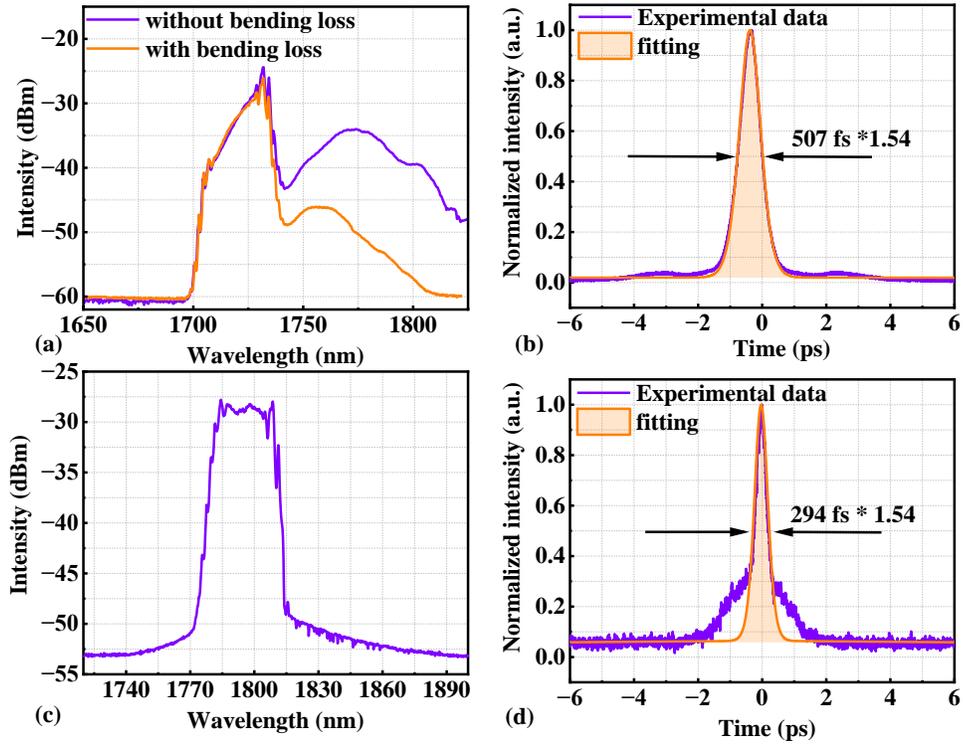

**Figure 3.** Spectra and autocorrelation traces after the Amplifier2 for 1720 nm (a,b) and 1800 nm (c,d).

### 3. Conclusions

In conclusion, we demonstrated a tunable CPA system operating from 1720 nm to 1800 nm and composed of only single-mode fibers. The Tm-doped fiber laser operated in a dissipative soliton regime, while intracavity tunability of the central wavelength was enabled by ATOF. The generated pulses were temporally stretched and consequently amplified in a two-cascade amplification system. When amplifying the pulse signal at a short wavelength, the strong ASE at the gain peak was effectively sup-pressed by 13 dB by introducing wavelength-dependent bending loss in the DCF. The tunable CPA system delivered 126 mW output power and 507 fs pulses at the shortest wavelength of 1720 nm. At the longest wavelength of 1800 nm, the maximum power level was 264 mW with 294 fs pulse duration. The wavelength tunability of high-power ultrafast lasers is highly desired in laser applications based on sensing or imaging techniques. With the merit of flexibility and compactness, the tunable CPA system is a strong rival among similar laser sources.

**Funding.** This research was funded by the European Union's Horizon 2020 research and innovation pro-gramme, grant number 871277, and by the Research Council of Finland, grant number 320165.